\documentclass[a4paper,12pt]{article}
 \pdfoutput=1

\usepackage[pdftex]{graphicx}
\usepackage{subfigure}
\usepackage{float}

\newcommand{\be}{\begin{equation}}
\newcommand{\ee}{\end{equation}}

\newcommand{\bea}{\begin{eqnarray}}
\newcommand{\eea}{\end{eqnarray}}

\newcommand{\p}{\partial}

\newcommand{\nn}{\nonumber \\}
\newcommand{\f}{\frac}
\newcommand{\w}{\wedge}
\newcommand{\ra}{\rightarrow}
\begin{document}
\thispagestyle{empty}
\begin{flushright}
{\bf arXiv:1712.09249}
\end{flushright}
\begin{center} \noindent \Large \bf Thermodynamics of 
Einstein-DBI System

\end{center}

\bigskip\bigskip\bigskip
\vskip 0.5cm
\begin{center}
{ \normalsize \bf   Shesansu Sekhar Pal
}

\vskip 0.5 cm

 Department of Physics, Utkal University,  Bhubaneswar, 751004, India
\vskip 0.5 cm
\sf { shesansu${\frame{\shortstack{AT}}}$gmail.com 
}
\end{center}
\centerline{\bf \small Abstract}

We continue our study of the Einstein-Dirac-Born-Infeld system as started in \cite{Pal:2012zn}. In this paper, we studied the thermodynamical aspect  of the   black hole by keeping the horizon  as planar in the grand canonical ensemble. 
In particular, we obtain the first law of thermodynamics and the Smarr type relation. Also, studied the  thermodynamic stability.



\newpage

\section{Introduction}

The study of AdS/CFT correspondence has opened up a new avenue to understand and interpret the gravitational solution that asymptotes to AdS at UV \cite{Maldacena:1997re}. In particular, the gravitational solution can predict the stability of the dual field theory.  Thermodynamically, from the gravitational point of view, one such simple criteria to examine is the  sign of   specific heat and the slope of the chemical potential with respect to charge density at fixed temperature, $\left(\f{\p\mu}{\p \rho}\right)_{T_H}$. If the system is unstable it my undergo  thermodynamic phase transition and go over to a stable phase. 

The thermodynamics  of charged black hole with spherical horizon that asymptotes to AdS spacetime is studied in greater detail  in the context of the Einstein-Maxwell system in \cite{Chamblin:1999tk}. One of the findings of the paper is that for geometry with such horizon  there exists a Hawking-Page phase transition. The details of the thermodynamics is calculated by regulating the divergent on-shell action via background subtraction method in \cite{Chamblin:1999tk}.

The singular nature of the gauge field in the Einstein-Maxwell system is well known and studied. It is solved in a seminal paper by Born-Infeld and Dirac (separately), by proposing an action (DBI) and its solution is constructed by demanding the {\it principle of finiteness} \cite{dbi}. 

The DBI action and its properties are well studied from point of view of string theory in \cite{Polchinski:1996na} and \cite{Myers:1999ps}. In fact, a lot has been studied by considering the DBI action as a probe brane, e.g., in \cite{Brattan:2012nb},   \cite{Bigazzi:2013jqa} and  \cite{Arean:2015wea}. In \cite{Giddings:2001yu} and  \cite{Mateos:2006yd} an attempt is made to understand the effect of the DBI action on the background geometry. In general it is very difficult to find the DBI-backreacted geometry of the spacetime  when the dimensionality of the DBI action is  less than that of the dimensionality of the background geometry. However, when both the dimensionalities are same, one can find  a solution.  The gravitational solution of the Einstein-Dirac-Born-Infeld action in generic dimension with space-filling brane and with  planar horizon is obtained in  \cite{Pal:2012zn}, as well as the dyonic solution in $3+1$ spacetime dimension. Different aspect of this  gravitational solution  is studied in \cite{Tarrio:2013tta} and \cite{Kundu:2017cfj}. However, the detailed thermodynamics of the system is yet to be studied\footnote{The thermodynamics of the Einstein-Born-Infeld system is studied for e.g., in \cite{Fernando:2006gh} and \cite{Myung:2008kd} and shown to exhibit  the Hawking-Page transition. Some properties of the Born-Infeld system is studied in \cite{Rasheed:1997ns}.  The phase diagram of the holographically constructed  Veneziano model is studied in \cite{Jarvinen:2011qe} }. Several applications of Einstein-DBI system in the context of AdS/CMT are studied in \cite{Pal:2012zn}, \cite{Charmousis:2010zz}, \cite{Kiritsis:2016cpm}, \cite{Blake:2017qgd}, \cite{Cremonini:2017qwq} and \cite{Lee:2011qu}.

 In this paper, we have studied the thermodynamics of the system by  adopting the counter  term method. Using such an approach, we have  calculated the free energy and other thermodynamical quantities. One of findings of the paper is to obtain the coefficient of the counter term that is required to generate a finite on-shell action at the boundary and determine its stability.


Upon inspection of   the black hole solution with planar horizon that are coming  from the Einstein-DBI system, we find   there exists  two  branches of the solution.  That is solutions with positive and negative tension of the DBI brane.


Demanding that the finite temperature gravitational solution that follows from such a system at the boundary gives us the AdS geometry provides us the following condition: $T_b+2\Lambda\equiv 2\Lambda_{eff} < 0$, where $T_b$  is the tension of the DBI action and $\Lambda$ is the background cosmological constant.
It just follows that the effective cosmological constant of the spacetime is not anymore just determined by $\Lambda$ alone but by the algebraic sum of the tension of the DBI action and the background cosmological constant. 
 Upon demanding that the gravitational solution respects the null energy condition allows us to keep only the positive tension of the brane. Since, $T_b$ is positive means that  $\Lambda$ can either be positive or negative, so that the effective cosmological constant can be negative,  $ 2 \Lambda_{eff}<0$.  In what follows, we shall assume that $\Lambda <0$. The negative value of the $T_b+2\Lambda$ suggests that the magnitude of the tension of the brane,  $T_b$, should be less than $2\Lambda$.   Moreover, the same condition, $T_b+2\Lambda\equiv 2\Lambda_{eff} < 0$, follows upon demanding that for large size of the horizon the Hawking temperature, $T_H$, and the energy, $E$, are positive.


We have shown that  the electrical charged black hole solution obeys the first law of thermodynamics. It is interesting to note that the first law of thermodynamics does not  depend on whether the solution respects  the null energy condition or not.
We have also studied the thermodynamics of the dyonic solution. 
For positive tension  of the DBI action, we find that either for an electrically or magnetically charged black hole or both there exists a minimum size of the horizon, which is above zero. The positivity of the Hawking temperature does not allow us to go below that minimum size. However, for a neutral black hole the size of the horizon can be made as small as we want.

The paper is organized as follows. In section two, we have described the system that we are interested in, namely the Einstein-DBI system and the solution that follows from it. The thermodynamic quantities like temperature and entropy just follows from the Hawking and Bekenstein-Hawking formula. The temperature of the black hole  is plotted versus the size of the horizon in fig(\ref{fig_1}). In fig(\ref{fig_2}), we have plotted the temperature of the black hole versus the entropy density. From the figure it just follows that  there exists two branches of the solution for negative tension of the DBI action. The specific heat  is negative for small size of the horizon whereas it is positive for larger size of  the horizon.  The null energy condition simply  rules out the possibility to have negative tension of the DBI action.

In section three, we have studied in detail the thermodynamics of the Einstein-DBI system. In particular, we have determined the precise coefficient of the counter term required to make the total action finite at the boundary. From the grand potential, we have calculated various thermodynamical quantities. We also show the existence of the first law of thermodynamics and the Smarr type formula. The thermodynamic stability is also studied.

In section four, we have calculated the expression of the energy density from the energy momentum tensor using the counter term method. We show that the expression of the energy density matches precisely with that follows from the thermodynamics.

In section five, we have studied the thermodynamics of the dyonic black hole. The details about the hypergeometric function is relegated to  Appendix A. In Appendix B, we have given the thermodynamic details of the dyonic solution that follows from  Einstein-Maxwell system  with planar horizon.

\section{The action}

The action that we consider is a bottom-up kind, which involves  different kind of  fields, namely,   the  metric,  abelian gauge field and  scalar field.  The matter field consists of  the gauge field and scalar field. The gauge field   has a non-linear from, namely the Dirac-Born-Infeld action \cite{dbi}, whereas the scalar field has a canonical  kinetic energy and a  potential energy term.  The precise form of the action that we are interested in  is
\be\label{eh_dbi_dilaton}
S=\f{1}{2\kappa^2}\int d^{d+1} x\bigg[ \sqrt{-g}\bigg({\cal R}-2\Lambda-\f{1}{2}\p_M\phi\p^M\phi-V(\phi)\bigg)-T_bZ_1(\phi)\sqrt{-det\bigg([g] Z_2(\phi)+\lambda F\bigg)_{MN}}\Bigg],
\ee 
where $[g]_{ab}=\p_aX^M\p_b X^N g_{MN}$ describes the  metric on  the world volume of the brane. For simplicity, we consider a space-filling brane. Using the   world volume of the brane diffeomorphism, we can fix the choice of the world volume coordinates with  the spacetime  coordinates. The indices  $M,~N$ etc. denote the spacetime index  and can take $d+1$ values.
$T_b$  and $\Lambda$ are  the tension of the brane, and background cosmological constant, respectively.  $F=dA$ is the two-form field strength that lives on the world volume of the brane. 

 It is easy to notice that the action as written down in  eq(\ref{eh_dbi_dilaton}) is in Einstein frame. The constant $\lambda$ is a dimension full object and has the dimension of length-squared  and in string theory it is identified with $\lambda=2\pi l^2_s$, where $l_s$ is the string length \cite{Myers:1999ps}. 
It is there to make the matrix $\left([g] Z_2(\phi)+\lambda F\right)_{MN}$ and its determinant dimensionless. 

\paragraph{EoM:} The equation of motion of the metric components that follows from it takes the following form
\bea\label{metric_eom}
&&{\cal R}_{MN}-\f{2\Lambda}{(d-1)}g_{MN}-\f{g_{MN}}{(d-1)}V(\phi)-\f{1}{2}\p_M\phi\p_N\phi-\nn&&\f{T_b~ Z_1(\phi)Z_2(\phi)}{4(d-1)}\f{\sqrt{-det\bigg(g~ Z_2(\phi)+\lambda F\bigg)_{PS}}}{\sqrt{-g}}\bigg[\bigg(g~ Z_2(\phi)+\lambda F\bigg)^{-1}+{\bigg(g~ Z_2(\phi)-\lambda F\bigg)}^{-1} \bigg]^{KL}\nn&&
\bigg[g_{MN}g_{KL}-(d-1)g_{MK}g_{NL} \bigg]=0,
\eea
where we have re-expressed the Einstein tensor in terms of the Ricci tensor, ${\cal R}_{MN}$, the energy momentum tensor and the trace of the energy momentum tensor. 

The gauge field equation of motion takes the following form
\be\label{gauge_field_eom}
\p_M\Bigg[Z_1(\phi)\sqrt{-det\bigg(g~ Z_2(\phi)+\lambda F\bigg)_{PS}}\Bigg(\bigg(g~ Z_2(\phi)+\lambda F\bigg)^{-1}-{\bigg(g~ Z_2(\phi)-\lambda F\bigg)}^{-1}  \Bigg)^{MN}\Bigg]=0
\ee

In what  follows,  the gauge field can be fully determined in terms of the metric components and the scalar field. Finally, the equation of motion of the scalar field
\bea\label{scalar_eom}
&&\p_{M}\bigg(\sqrt{-g}\p^M\phi \bigg)-\sqrt{-g}\f{dV(\phi)}{d\phi}-T_b \f{dZ_1(\phi)}{d\phi}\sqrt{-det\bigg(g~ Z_2(\phi)+\lambda F\bigg)_{KL}}-\nn&&\f{T_b}{2}Z_1(\phi)
\sqrt{-det\bigg(g~ Z_2(\phi)+\lambda F\bigg)_{KL}}\f{dZ_2(\phi)}{d\phi}{\bigg(g~ Z_2(\phi)+\lambda F\bigg)}^{-1MN}g_{MN}=0.
\eea

\paragraph {Ansatz:} For our purpose, we shall consider an ansatz where the metric is diagonal and there exists a rotational symmetry along the (d-1) plane involving the $ x_i$'s. The abelian field strength and the scalar field is assumed to be of the following form 
\be\label{ansatz_solution}
ds^2_{d+1}=-g_{tt}(r)dt^2+g_{rr}(r)dr^2+g_{xx}(r)dx^2_i,\quad
A=A_t(r)dt,\quad F=A_t' dr\w dt,\quad \phi={\rm Constant},
\ee 
essentially, the fields are considered to be function of the radial coordinate only. It means we do have  a rotational symmetry in the $x_i$ directions. For general solution with non-trivial value of the scalar field at IR can be found in \cite{Pal:2012zn}.

\paragraph{UV:} Using  the above mentioned coordinate system,  the radial coordinate is denoted as, $r$.  The UV is defined to be at $r\ra \infty$. In which case, we assume  the spacetime asymptotes to AdS spacetime. Moreover, the gauge field strength becomes non-singular for trivial scalar field. The black hole solution for $Z_1=1=Z_2$ reads as \cite{Pal:2012zn}
\bea\label{sol_uv}
ds^2_{d+1}&=&\f{r^2}{R^2}\left[-f(r)dt^2+dx^2_i\right]+\f{R^2 dr^2}{r^2 f(r)}, ~~~~~\lambda A'_t=\f{\rho R^{d-1}}{\sqrt{\rho^2 R^{2(d-1)}+r^{2(d-1)}}}\nn
f(r)&=&\f{c_1}{r^d}-\f{2\Lambda R^2}{d(d-1)}-\f{T_b R^2\rho}{(d-1)}\f{r^{1-d}}{R^{1-d}}~~~{}_2F_1\bigg[-\f{1}{2},\f{1}{2(d-1)},\f{2d-1}{2(d-1)},-\f{r^{2(d-1)}}{R^{2(d-1)}\rho^2}\bigg],
\eea
where $R,~\rho$ are the radius of the AdS spacetime and the charge density, respectively. $c_1$ is the constant of integration and identified with the mass of the black hole.

\paragraph{ Temperature, Entropy and Chemical potential:} Some of the salient features of such black hole solution is as follows. 
The Hawking temperature, the Bekenstein- Hawking entropy density  and the chemical potential, $\mu$, associated to such a black hole reads as
\bea\label{temp}
T_H&=&-\f{ r_h}{(d-1)4\pi }\Bigg[2\Lambda+T_b~ r^{1-d}_h\sqrt{\rho^2R^{2(d-1)}+r^{2(d-1)}_h}\Bigg],~s=\f{2\pi}{\kappa^2}\Bigg( \f{r_h}{R}\Bigg)^{d-1}\nn
\mu&=&-\f{r_h}{\lambda} ~{}_2F_1\Bigg[\f{1}{2},\f{1}{2(d-1)},\f{1-2d}{2-2d},-\f{ r^{2(d-1)}_h}{\rho^2R^{2(d-1)}}\Bigg],
\eea
where  ${}_2F_1[a,b,c,z]$ is the hypergeometric function. Note the chemical potential is defined as the value of the gauge potential evaluated at the boundary, $\mu=A_t(\infty)-A_t(r_h)$, where $r_h$ is the horizon and $A_t(r_h)=0$. 

Let us construct few dimensionless quantities, $x_h\equiv r_h/R,~t_b\equiv T_b R^2,~{\tilde \lambda}\equiv\Lambda R^2,~t_H\equiv T_HR$, in which case, we can rewrite the temperature, chemical potential and entropy density as
\bea\label{temp_chem}
t_H&=&-\f{ x_h}{(d-1)4\pi }\Bigg[2{\tilde \lambda}+t_b~ x^{1-d}_h\sqrt{\rho^2+x^{2(d-1)}_h}\Bigg],\nn
\f{\mu}{R}\lambda&=&-x_h ~{}_2F_1\Bigg[\f{1}{2},\f{1}{2(d-1)},\f{1-2d}{2-2d},-\f{ x^{2(d-1)}_h}{\rho^2}\Bigg],\quad
s=\left(\f{2\pi}{\kappa^2}\right)x^{d-1}_h=\f{x^{d-1}_h}{4G_N},
\eea

\begin{figure}[h!]
\centering
   {\includegraphics[ width=8cm,height=6cm]{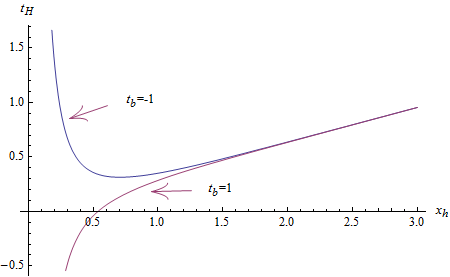} }
  \caption{
   The  figure is   plotted for dimensionless Hawking  temperature, $t_H$, vs the size of the horizon, $x_h$, for  $AdS_5$ black hole.  The parameters are set as: ${\tilde \lambda}=-\f{d(d-1)}{2}-\f{t_b}{2},~d=4,~\rho=-2$. For negative brane tension there exists two branches whereas for positive brane tension there exists only one branch.  }
 \label{fig_1}
\end{figure}

The dimensionless temperature is plotted versus the size of the horizon in fig(\ref{fig_1}) for fixed charge density as well as for $d=4$. Note, the temperature is an even function in the charge density, $\rho$. So, as far as the temperature and chemical potential are considered, the positive,  or the negative  charge density,   has the same effect for a fixed horizon.

\begin{figure}[h!]\label{fig_2}
\centering
   {\includegraphics[ width=8cm,height=6cm]{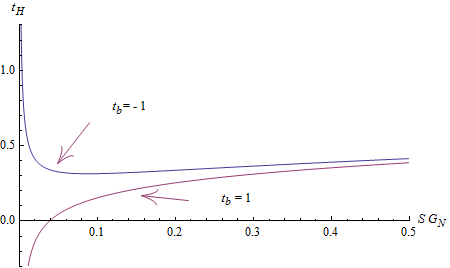} }
  \caption{
   The  figure is   plotted for the temperature vs the entropy density, $s$, times the Newton constant, $G_N$,  of the  $AdS_5$ black hole.  The parameters are set as: ${\tilde \lambda}=-\f{d(d-1)}{2}-\f{t_b}{2},~d=4,~\rho=-2$. For negative brane tension there exists two branches whereas for positive brane tension there exists only one branch.  }
\end{figure}

Finally, we can express the dimensionless temperature, $t_H$,  in terms of the entropy density, $s$, as 
\be
t_H=-\f{(4G_N s)^{\f{1}{d-1}}}{4\pi(d-1)}\left[ 2{\tilde \lambda}+\f{t_b}{4G_N s}\sqrt{\rho^2+(4G_N s)^2}\right]
\ee
 and is plotted in fig(\ref{fig_2}).
 
\paragraph{Temperature versus entropy density or horizon:} If we look at fig(\ref{fig_1}) and  fig(\ref{fig_2}), then it is easy to notice that the temperature decreases for small values of the  horizon size (or entropy density)  and increases for higher values of the  horizon size  (or entropy density) when the tension of the brane is positive.  

If we  define  the specific heat for fixed charge density, $C_{\rho}$,  as, $C_{\rho}=T_H \left(\f{\p s}{\p T_H}\right)_{\rho}$ 
then it follows from fig(\ref{fig_1}) and fig(\ref{fig_2}) that there exists  branches  for which  $C_{\rho}$ changes sign. For negative, $t_b$, and for small entropy density the specific heat is negative whereas for large entropy density it is positive. In the former case the system is thermodynamically unstable whereas in the latter case it is stable.

\paragraph{Chemical potential versus  horizon:} The dimensionless chemical potential as written down in eq(\ref{temp_chem}) does not  depend explicitly  on the sign of the tension of the DBI action, $T_b$, and is independent of the sign of the charge density. For fixed non-zero charge density, $\rho$, it starts from zero and saturates to some finite negative value for large size of the horizon. More importantly, it never diverges. The slope of the curve between chemical potential and the horizon for fixed charge density is negative and approaches zero for large size of the horizon.

The behavior of the chemical potential that follows from Einstein-Maxwell system is different in many respects, e.g., it diverges for very small size of the horizon, vanishes for large size of the horizon. The detailed comparison at UV and at IR is given in the conclusion.

\paragraph{Null energy condition:}

 It is expected that the gravitational solution should respect the null energy condition: $T_{MN} u^M u^N \geq 0$, for some null vectors $u^M$ and energy momentum tensor $T^{MN}$. Using the  equation for the metric tensor, we can rewrite the null energy condition as ${\cal R}_{MN} u^M u^N \geq 0$, where ${\cal R}_{MN}$ is the Ricci tensor. 

Given the solution as written down in eq(\ref{ansatz_solution}), we can make two different choices of the null unit vectors. These are (a) $u^t=1/\sqrt{g_{tt}},~~ u^r=1/\sqrt{g_{tt}},~~u^{x_i}=0$ and the second choice is (b)  $u^t=1/\sqrt{g_{tt}},~~ u^r=0,~~u^{x_1}=1/\sqrt{g_{xx}}$ and $u^{x_i}=0$ for $i\neq 1$. The choice (a)  saturates the null energy condition and hence does not give anything interesting. However choice (b) gives 
\be
{\cal R}_{MN} u^M u^N=\f{r^2f''(r)+(d+1)r f'(r)}{2R^2}=\f{T_b}{2} \left(\f{R}{r} \right)^{d-1}\f{\rho^2}{\sqrt{\rho^2+\left(\f{r}{R} \right)^{2(d-1)}}},
\ee

where we have used the solution as written explicitly in eq(\ref{sol_uv}). 
Finally, the imposition of the null energy condition on the solution eq(\ref{sol_uv}) gives the positive tension of the brane, $T_b \geq 0$.

\section{Thermodynamics}

In this section, we would like to study the thermodynamics of the Einstein-DBI-dilaton  system for trivial value of the scalar field. It means we are looking at the  thermodynamics at UV, where the spacetime approaches AdS. The approach that we shall follow to study thermodynamics is that of the covariant counter term method. 

The   on-shell bulk gravitational  action eq(\ref{eh_dbi_dilaton})  becomes
\be
S_{bulk}=-\f{1}{2\kappa^2}\int d^{d+1}x \f{d}{dr}\left(\f{\sqrt{g_{tt}}}{\sqrt{g_{rr}}}g\rq{}_{xx}g^{\f{(d-3)}{2}}_{xx} \right), \quad {\rm where}\quad  g\rq{}_{xx}=\f{d}{dr} (g_{xx})
\ee

and the $(d+1)$ dimensional  volume integral reduces to a surface integral: \\$S_{bulk}=-\f{1}{2\kappa^2}\int d^{d}x \left(\f{\sqrt{g_{tt}}}{\sqrt{g_{rr}}}g\rq{}_{xx}g^{\f{(d-3)}{2}}_{xx} \right)^{UV}_{IR} $, where UV is defined at $r=\infty$ and the IR at $r=r_h$. The Gibbons-Hawking term required for the proper variation of the action eq(\ref{eh_dbi_dilaton}) 
\be\label{gh_action}
S_{GH}=-\f{1}{\kappa^2}\int d^d x\sqrt{-h} K,
\ee
where $K$ is the trace of the extrinsic curvature and $h_{ab}$ is the induced metric on the space like/time like surface on the boundary. The extrinsic curvature is defined as $K_{ab}=-\f{1}{2}(\nabla_{a}n_{b}+\nabla_{b}n_{a})$, where $n_{a}$ is the unit vector normal to the boundary.
The counter term that is required to find the well defined on-shell bulk action on the boundary 
\be\label{ct_action}
S_{ct}=\f{\alpha}{2\kappa^2}\int d^d x \sqrt{-h},
\ee
where $\alpha$ is a  constant to be  determined by demanding that $S_{bulk}+S_{GH}+S_{ct}$ is well defined and finite on the boundary.

Using the solution as written in eq(\ref{ansatz_solution}), we find the form of $S_{GH}$ and $S_{ct}$ as 
\be
S_{GH}=\f{1}{2\kappa^2}\int d^d x\left(\f{g\rq{}_{tt}g^{\f{(d-1)}{2}}_{xx}}{\sqrt{g_{tt}g_{rr}}} +(d-1)g\rq{}_{xx}\f{\sqrt{g_{tt}}}{\sqrt{g_{rr}}}g^{\f{(d-3)}{2}}_{xx}\right),\quad S_{ct}=\f{\alpha}{2\kappa^2}\int d^d x \sqrt{g_{tt}}g^{\f{(d-1)}{2}}_{xx}.
\ee

Since, we are interested to understand the thermodynamics of the  DBI system at UV, we shall consider the solution as written in eq(\ref{ansatz_solution}) for this purpose. The sum of all the three terms, $S_{total}\equiv S_{bulk}+S_{GH}+S_{ct}$ to be evaluated on the boundary is
\bea
S_{total}&=&\f{1}{2\kappa^2}\int d^{d}x \Bigg[\left(-\f{\sqrt{g_{tt}}}{\sqrt{g_{rr}}}g\rq{}_{xx}g^{\f{(d-3)}{2}}_{xx}\right)^{UV}_{IR}\nn &+&\left(\f{g\rq{}_{tt}g^{\f{(d-1)}{2}}_{xx}}{\sqrt{g_{tt}g_{rr}}} +(d-1)g\rq{}_{xx}\f{\sqrt{g_{tt}}}{\sqrt{g_{rr}}}g^{\f{(d-3)}{2}}_{xx}+\alpha  \sqrt{g_{tt}}g^{\f{(d-1)}{2}}_{xx}\right)^{UV}\Bigg]
\eea

In order to evaluate  $S_{total}$, let us consider  the precise form of the geometry as written in eq(\ref{sol_uv}), in which case, it reduces to 
\be
S_{total}=\f{1}{2\kappa^2}\int d^{d}x\Bigg( \left[ \f{r^d}{R^{d+1}}\left(r~f\rq{}(r)+2(d-1)f(r)+\alpha R\sqrt{f(r)}\right)\right]_{UV}+2\left( \f{r^d}{R^{d+1}} f(r)\right)_{IR}\Bigg).
\ee

Note, the function, $f(r)$, obeys the following differential equation
\be
rf\rq{}(r)+df(r)+\f{2\Lambda R^2}{d-1}+\f{T_bR^2}{d-1} \f{r^{1-d}}{R^{1-d}}\sqrt{\rho^2+\f{r^{2(d-1)}}{R^{2(d-1)}}}=0
\ee

and the function $f(r)$ vanishes at IR, i.e., $f(r_{IR})=0$. In which case

\be\label{S_total}
S_{total}=\f{V_{d-1}\beta}{2\kappa^2}\Bigg(\f{r^d}{R^{d+1}} \left[-\f{2\Lambda R^2}{d-1}-\f{T_bR^2}{d-1} \f{r^{1-d}}{R^{1-d}}\sqrt{\rho^2+\f{r^{2(d-1)}}{R^{2(d-1)}}}+(d-2)f(r)+\alpha R\sqrt{f(r)}\right]\Bigg)^{UV},
\ee

where $\int d^{d}x=\int d^{d-1}x\int dt=V_{d-1}~\beta$ and $\beta=1/T_H$.  The time direction is considered to be periodic with periodicity $\beta$, which is  the inverse of Hawking temperature. Recall, the background cosmological constant, $\Lambda$, is negative and the function, $f(r)$, has the form given as
\be
f(r)=\f{c_1}{r^d}-\f{2\Lambda R^2}{d(d-1)}-\f{T_b R^2\rho}{(d-1)}\f{r^{1-d}}{R^{1-d}}~~~{}_2F_1\bigg[-\f{1}{2},\f{1}{2(d-1)},\f{2d-1}{2(d-1)},-\f{r^{2(d-1)}}{R^{2(d-1)}\rho^2}\bigg]. 
\ee

The asymptotic behavior of the quantity written in the square bracket of eq(\ref{S_total}) is given in the appendix A, using it we find at UV, the on-shell value of the action  diverges unless we set 
\be\label{alpha}
\alpha=-2\f{(d-1)}{R}\sqrt{\f{-2\Lambda_{eff}R^2 }{d(d-1)}}.
\ee
For this choice of, $\alpha$, we get the regularized on-shell action,  $S_{total}$, as
\be
S_{total}=-\f{V_{d-1}\beta}{2\kappa^2}\f{c_1}{R^{d+1}}\equiv -I,
\ee
where $I/\beta$ is the thermodynamic potential in the grand canonical ensemble. 
The quantity, $c_1$ is determined from the condition $f(r_{IR})=0$.  In case of a black hole, $r_{IR}$, is the location of the horizon, $r_h$. Finally
\bea
I&=&\f{V_{d-1}\beta}{2\kappa^2}\Bigg(\f{2\Lambda }{d(d-1)} \f{r^d_h}{R^{d-1}}+\f{T_b \rho}{(d-1)}r_h~~~{}_2F_1\bigg[-\f{1}{2},\f{1}{2(d-1)},\f{2d-1}{2(d-1)},-\f{r^{2(d-1)}_h}{R^{2(d-1)}\rho^2}\bigg]\Bigg).
\eea


\paragraph{Entropy:} Various thermodynamic quantities are determined from the thermodynamic potential, $I/\beta$, by differentiation. As an example, the entropy   is
\be
S=\beta\left(\f{\p I}{\p \beta}\right)_{\varphi}-I
\equiv \f{M}{N},
\ee
where $(\f{\p I}{\p \beta})_{\varphi}$ is to be evaluated at constant chemical potential. The quantities $M$ and $N$ are
\bea
M&=&\left(\f{ r_h}{R}\right)^{d-1} V_{d-1} \sqrt{\rho^2+r^{2 (d-1)}_h  R^{2 (1-d)}} ~
 \Bigg(2 R^{d+1} r^{d+ 1}_h \Lambda \rho +  d R^{2 d}  r^2_h T_b\rho^2\nn&& {}_2F_1\Bigg[-\f{1}{2}, \f{1}{ 2( d-1)}, 1 + \f{1}{ 2( d-1)}, -\f{r^{2(d-1)}_h}{R^{2(d-1)}\rho^2}\bigg]  - 2 R^{ d+1} r^{d+1}_h \Lambda \sqrt{\rho^2+r^{2 (d-1)}_h  R^{2 (1-d)}}\nn&&
  {}_2F_1\Bigg[\f{1}{2},  \f{1}{ 2( d-1)}, 1 + \f{1}{ 2( d-1)}, -\f{r^{2(d-1)}_h}{R^{2(d-1)}\rho^2}\bigg]  -   T_b r^{2}_hR^{2d}
 ( \rho^2+r^{2 (d-1)}_h  R^{2 (1-d)})\nn&& 
   {}_2F_1\Bigg[\f{1}{2}, \f{1}{ 2( d-1)}, 1 + \f{1}{ 2( d-1)}, -\f{r^{2(d-1)}_h}{R^{2(d-1)}\rho^2}\bigg] \Bigg),\nn
N&=&  4 G\rho \Bigg(T_bR^{2d} r^{2}_h   (\rho^2 +  r^{2 (d-1)}_h  R^{2 (1-d)}) + 
   2   \Lambda R^{ d+1} r^{d+1}_h  \sqrt{\rho^2+r^{2 (d-1)}_h  R^{2 (1-d)}}+\nn&&
   \f{1}{\rho}(\rho^2+r^{2 (d-1)}_h  R^{2 (1-d)} )  {}_2F_1\Bigg[1,  \f{d}{ 2( d-1)}, 1 + \f{1}{ 2( d-1)}, -\f{r^{2(d-1)}_h}{R^{2(d-1)}\rho^2}\bigg] \times\nn&&
   \bigg  ((d-2) R^{2 d}    r^2_h T_b  \rho^2 -
      T_b R^2 r^{2d}_h -2   \Lambda R^{ d+1} r^{d+1}_h  \sqrt{\rho^2+r^{2 (d-1)}_h  R^{2 (1-d)}}\bigg)\Bigg) 
\eea

 In general,  it  is difficult to simplify further the ratio, $M/N$. So, we shall find it for large value of the charge density, in which case, the entropy density reduces to
 \be
 s\equiv\f{S}{V_{d-1}}=\f{r^{d-1}_h}{4GR^{d-1}}+{\cal O}(1/\rho^{22}),\quad 2\kappa^2\equiv16\pi G.
 \ee
Even though, we did not find an exact result, but this holds true for very high order in $1/\rho$.
In fact, the same value of the entropy density, $s=\f{r^{d-1}_h}{4GR^{d-1}}$, follows  for  $d=4$, to very high order, for  the  small and large  horizon size.

\paragraph{Chemical potential:} The chemical potential, $\mu$, is defined as the difference of the electrostatic potential at the boundary and at the horizon, i.e., $\mu=A_t(r=\infty)-A_t(r=r_h)$. 
In our case,   the gauge potential  has the following form
\be
A_t(r)=\f{r}{\lambda} ~{}_2F_1\Bigg[\f{1}{2},\f{1}{2(d-1)},\f{1-2d}{2-2d},-\f{r^{2(d-1)}}{R^{2(d-1)}\rho^2}\Bigg]+\varphi,
\ee
where the constant, $\varphi$, is defined as  the location for which, $A_t(r_h)=0$. 
Using the asymptotic behavior of the hypergeometric function as discussed in Appendix A, we find that $A_t(r)|_{r=boundary} \sim \f{1}{\lambda r^{d-2}}+\varphi$. So, the gauge potential becomes constant  at the boundary whereas the  electric field vanishes at the boundary.
It means the chemical potential 
\be
\mu=\varphi=-\f{r_h}{\lambda} ~{}_2F_1\Bigg[\f{1}{2},\f{1}{2(d-1)},\f{1-2d}{2-2d},-\f{r^{2(d-1)}_h}{R^{2(d-1)}\rho^2}\Bigg].
\ee

The electric charge, $q$,  is defined from thermodynamics point of view as
\be\label{charge}
q=-\f{1}{\beta}\left(\f{\p I}{\p \varphi}\right)_{\beta}=\f{T_b V_{d-1} \lambda}{16\pi G}\rho+{\cal O}(1/\rho^{12}).
\ee

 
\paragraph{Energy:} Thermodynamically, the energy is defined as 
\be
E=\left(\f{\p I}{\p \beta}\right)_{\varphi}-\f{\varphi}{\beta}\left(\f{\p I}{\p \varphi}\right)_{\beta}
\ee

This is due to the fact that the thermodynamic potential is defined as $I/\beta=(E-T_H S-\varphi q)$. Finally,  the energy came out as
\bea
E&=&-\f{(d-1)V_{d-1}}{16\pi G}~\f{c_1}{R^{d+1}}=-\f{(d-1)V_{d-1}}{16\pi G}\times\nn&& \Bigg(\f{2\Lambda }{d(d-1)} \f{r^d_h}{R^{d-1}}+\f{T_b \sqrt{\rho^2}}{(d-1)}r_h~~~{}_2F_1\bigg[-\f{1}{2},\f{1}{2(d-1)},\f{2d-1}{2(d-1)},-\f{r^{2(d-1)}_h}{R^{2(d-1)}\rho^2}\bigg]\Bigg).
\eea

\begin{figure}[h!]\label{fig_3}
\centering
   {\includegraphics[ width=8cm,height=6cm]{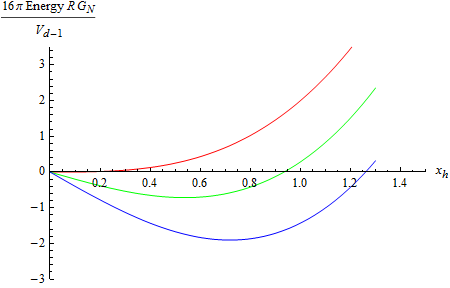} }
  \caption{
   The  figure is   plotted for the   energy density times $16\pi G_N R$ vs the  horizon size for the  $AdS_4$ black hole.  The parameters are set as: ${\tilde \lambda}=-\f{d(d-1)}{2}-\f{t_b}{2},~\rho=-2$. The upper curve, middle curve and the lower curves are for $t_b=0,~~t_b=1$ and $t_b=2$, respectively.}
\end{figure}

Again this result holds for very high order in $1/\rho$.
The energy density is plotted  for ($\rho=- 2$)   negative charge density in fig(\ref{fig_3}) . For very high value of the size of the horizon, $r_h\ra \infty$, the energy, $E\ra -\f{(T_b+2\Lambda )}{d R^{d-1}} r^d_h \f{V_{d-1}}{16\pi G_N}$. If we demand that the energy should become positive then it gives us the condition $T_b+2\Lambda <0$.


\paragraph{First law of thermodynamics:} It is known that the first law of thermodynamics, in the presence of chemical potential (electric type) takes the following form
\be
dE=T_{H}~dS+\varphi~dq.
\ee

Upon performing explicit calculations, we find the variation of the energy reads as
\bea
dE&=&-\f{V_{d-1}}{16\pi G}\left(\f{2\Lambda}{R^{d-1}} r^{d-1}_h+T_d~\rho\sqrt{1+\f{r^{2(d-1)}_h}{R^{2(d-1)}\rho^2}} \right)dr_h+T_b\f{V_{d-1}}{16\pi G} \f{r_h}{(d-1)}\nn&&\left(\sqrt{1+\f{r^{2(d-1)}_h}{R^{2(d-1)}\rho^2}}-d~ {}_2F_1\Bigg[-\f{1}{2},\f{1}{2(d-1)},1+\f{1}{2(d-1)},-\f{r^{2(d-1)}_h}{R^{2(d-1)}\rho^2}\Bigg]\right)d\rho
\eea

Now, using the identity
\bea
&&\sqrt{1+\f{r^{2(d-1)}_h}{R^{2(d-1)}\rho^2}}-d~ {}_2F_1\Bigg[-\f{1}{2},\f{1}{2(d-1)},1+\f{1}{2(d-1)},-\f{r^{2(d-1)}_h}{R^{2(d-1)}\rho^2}\Bigg]\nn&=&-(d-1)~ {}_2F_1\Bigg[\f{1}{2},\f{1}{2(d-1)},1+\f{1}{2(d-1)},-\f{r^{2(d-1)}_h}{R^{2(d-1)}\rho^2}\Bigg]
\eea

we ended up
\bea\label{1st_law_temp_electric_charge}
dE&=&-\f{V_{d-1}}{16\pi G}\left(\f{2\Lambda}{R^{d-1}} r^{d-1}_h+T_b~\rho\sqrt{1+\f{r^{2(d-1)}_h}{R^{2(d-1)}\rho^2}} \right)dr_h-T_b\f{V_{d-1}}{16\pi G} r_h\nn&&{}_2F_1\Bigg[\f{1}{2},\f{1}{2(d-1)},1+\f{1}{2(d-1)},-\f{r^{2(d-1)}_h}{R^{2(d-1)}\rho^2}\Bigg]d\rho=T_H~dS+\varphi~dq,
\eea

where we have used the expression of the temperature, the chemical potential and the charge from eq(\ref{temp}) and  eq(\ref{charge}).

It is interesting to note that the validity of the first law of thermodynamics is independent of the sign of the tension of the DBI action. This suggests that the first law of thermodynamics does not depend on the validity of the null energy condition.

\paragraph{Smarr type formula:} Given the energy, temperature, entropy, chemical potential and charge, let us find a Smarr type equation involving these quantities. It is easy to obtain the following relation by explicit calculations

\be\label{smarr_relation}
E=\f{(d-1)}{d}\left(T_H~S+\varphi~ q\right).
\ee

\paragraph{Modified Gibbs-Duhem relation:} In thermodynamics,  the Gibbs-Duhem equation relates the differentials of the intensive variables.  However,  in the studies of the charged black holes that  asymptotes to AdS spacetime  satisfies the following modified relation involving the change in temperature to the change in the chemical potential with  the change in energy
\be
\f{(d-1)}{d}\left(s~~ dT_H+q~~d\varphi \right)=d E
\ee

This follows from the first law of thermodynamics and the Smarr equation as written in  eq(\ref{1st_law_temp_electric_charge}) and eq(\ref{smarr_relation}), respectively. 
It should be noted that  we have the differentials of the intensive variables like temperature, $dT_H$,  and chemical potential, $d\varphi$, that  appear in the left hand side of the equation whereas on the right hand side we have the differential of the  extensive variable like energy, $dE$.

\paragraph{ Stability: } It is suggested in \cite{Faedo:2017aoe} that the thermodynamical stability in grand canonical ensemble can be studied by looking at the positivity of the following quantities
\be
 T_H\left(\f{\p s}{\p T_H}\right)_{\rho} >0; \qquad \chi^{-1}\equiv\left(\f{\p\mu}{\p \rho}\right)_{T_H} >0
\ee
where $\mu, ~\rho, ~s, ~T_H$ are  chemical potential, charge density, entropy density and Hawking temperature, respectively. 

The ratio of the specific heat with the entropy density,  $\f{C_{\rho}}{s}$, which is a dimensionless quantity, gives an idea of the stability of the system. The specific heat is defined as $C_{\rho}\equiv T_H\left(\f{\p s}{\p T_H}\right)_{\rho}$. Hence, the quantity, $\f{\p(log~s)}{\p(log~T_H)}=\f{C_{\rho}}{s}$. Upon calculating it explicitly

\be
\f{C_{\rho}}{s}=(d-1)\left(1+\f{T_b \rho^2 R^{2(d-1)}}{4\pi T_H r^{d-2}_h\sqrt{\rho^2 R^{2(d-1)}+r^{2(d-1)}_h}} \right)^{-1}.
\ee

It follows very easily that the sign of  $\f{C_{\rho}}{s}$ is completely determined by the sign of  $T_b$. If $T_b$ is considered to be positive then it means the system that we are considering is thermodynamically stable. However, if we choose $T_b$  to be negative then the system can have negative $\f{C_{\rho}}{s}$, which means that  it can be thermodynamically unstable. 
 Moreover, in the negative $T_b$ case the system can undergo thermodynamic phase transition because of the divergence in  specific heat. However, negative $T_b$ has been ruled out by the null energy condition.

\begin{figure}[h!]
	\centering
	{\includegraphics[ width=8cm,height=6cm]{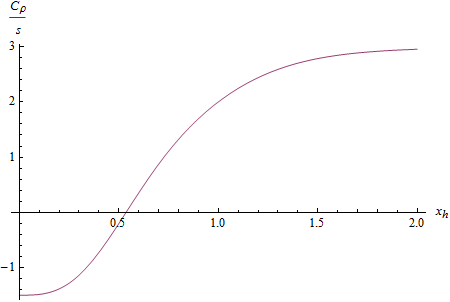} }
	\caption{
		The  figure is   plotted for the   specific  heat over entropy density  versus the  horizon size for the  $AdS_5$ black hole.  The parameters are set as: ${\tilde \lambda}=-\f{d(d-1)}{2}-\f{t_b}{2},~\rho=-2, t_b=1$ .}
	\label{fig_4}
\end{figure}

If we make a plot for ratio of the specific heat over the entropy density, $\frac{C_{\rho}}{s}$, for the positive tension of the brane, we find that for very small size of the horizon the above mentioned ratio  becomes negative, see fig(\ref{fig_4}). This is not surprising because for such a value of the size of the horizon, the temperature itself becomes negative. It means there exists a minimum size of the horizon for charged black hole.  In fact, this behavior is peculiar to the charged AdS black hole, see fig(\ref{fig_8}) of the Einstein-Maxwell system in Appendix B. 

For both positive as well as negative tension of the DBI brane, the ratio $\f{C_{\rho}}{s}$ approaches $(d-1)$ for large size of the horizon (which is same as large black holes). However, when the size of the black hole becomes small in units of the size of the AdS radius, the ratio $\f{C_{\rho}}{s}$ can be negative. When the tension of the DBI brane is positive one cannot go below a minimum size of the horizon, which is determined by the positive value  of the Hawking temperature. 

\begin{figure}[h!]
	\centering
	{\includegraphics[ width=8cm,height=6cm]{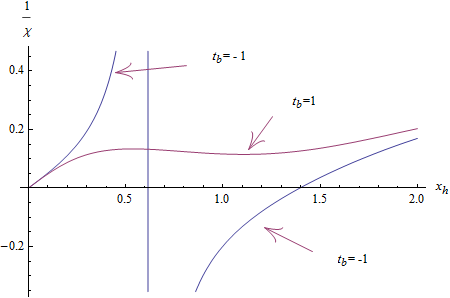} }
	\caption{
		The  figure is   plotted for the  change in chemical potential with the charge density for a fixed temperature   versus the  horizon size for the  $AdS_5$ black hole.  The parameters are set as: ${\tilde \lambda}=-\f{d(d-1)}{2}-\f{t_b}{2},~\rho=-2, t_b=1$ .}
	\label{fig_5}
\end{figure}

As far as the other quantity, $\chi^{-1}>0$, is concerned,  it is very difficult to find its sign, analytically. However,  one can read out the sign from the plot of such a quantity, $1/\chi$, versus the size of the horizon for fixed charge density, $\rho$, which is  drawn in fig(\ref{fig_5}). It follows that $\chi^{-1}>0$ can become negative for negative tension of the brane. But such a value of the tension of the brane has been ruled out by the null energy condition.

So, the gravitational solution under study is thermodynamically stable.

\section{Energy momentum tensor}

The energy  momentum tensor for this case has the form
\be
T_{ab}=\f{1}{8\pi G}\left(K_{ab}-Kh_{ab}+\f{\alpha}{2}h_{ab} \right)
\ee

The energy is calculated as follows
\be
E=\int d^{d-1}x \sqrt{det(\sigma)}~N~ u^t~ u^t~ T_{tt},
\ee
where the geometry on the boundary is written in the ADM way  as 
\be
h_{ab}dx^adx^b=-N^2 dt^2+\sigma_{ij}(dx^i+N^i dt)(dx^j+N^j dt)
\ee
and $u^t$ is the unit timelike vector normal to the boundary. Note, the  integrand in the calculation of the energy  is to be evaluated on the boundary and the integration is to be done over the spatial directions in field theory. The energy takes the following form
\be
E=\int d^{d-1}x~ g^{\f{d-1}{2}}_{xx} \f{1}{\sqrt{g_{tt}}} \f{\left[K_{tt}+K g_{tt}-\f{\alpha}{2}g_{tt} \right]}{8\pi G},~ u^t=\f{1}{\sqrt{g_{tt}}},~ \sqrt{det(\sigma)}=g^{\f{d-1}{2}}_{xx},~N=\sqrt{g_{tt}} 
\ee

Calculating   the extrinsic curvatures  we find 
\be
E=-\f{1}{8\pi G}\int d^{d-1}x~~~~ g^{\f{d-1}{2}}_{xx}\left[\f{(d-1)g\rq{}_{xx}g_{tt}}{2g_{xx}}+\f{\alpha}{2}\sqrt{g_{tt}}\right]
\ee

By considering the metric components as $g_{tt}=\f{r^2}{R^2} f(r),~g_{xx}= \f{r^2}{R^2}$, we find the energy reduces to 
\be
E=-\f{V_{d-1}}{8\pi G}\f{r^d}{R^d}\left[\f{(d-1)}{R}f(r)+\f{\alpha}{2}\sqrt{f(r)} \right],\quad \int d^{d-1}x=V_{d-1}
\ee

Let the function, $f(r)$, at the boundary goes as 
\be
lim_{r\ra \infty} f(r)=c_0+\f{c_1}{r^d}+{\cal O} (\f{1}{r^{d+1}}),
\ee

In which case the energy at the boundary becomes 
\be\label{energy}
E=-\f{(d-1)}{16\pi G}\f{V_{d-1}}{R^{d+1}} c_1,
\ee
where we have  used $\alpha=-2(d-1)\sqrt{c_0}/R$. For electrically charged black hole, $c_0=-\f{2\Lambda_{eff} R^2}{d(d-1)}$, see Appendix A, in which case, $\alpha$ matches precisely with that written in eq(\ref{alpha}).

\section{Dyonic  Black hole }

The dyonic solution for the Einstein-Hilbert-DBI system is found in \cite{Pal:2012zn}. Let us write down  the solution to the geometry as well as to the gauge field strength
\bea\label{dyon_sol}
ds^2_{3+1}&=&\f{r^2}{R^2}\left[-f(r)dt^2+dx^2+dy^2\right]+\f{R^2 dr^2}{r^2 f(r)},\quad F=A\rq{}_t(r)dt\w dr+B dx\w dy\nn
A_t(r)&=&\f{r\rho}{\lambda\sqrt{\rho^2+\lambda^2 B^2}}{}_2F_1\bigg[\f{1}{2},\f{1}{4},\f{5}{4},-\f{r^{4}}{R^4(\rho^2+\lambda^2 B^2)}\bigg]+\varphi_E,
\eea
where $B$ and $\varphi_E=-\f{r_h \rho}{\lambda\sqrt{\rho^2+\lambda^2 B^2}}{}_2F_1\bigg[\f{1}{2},\f{1}{4},\f{5}{4},-\f{r^{4}_h}{R^4(\rho^2+\lambda^2 B^2)}\bigg]$ are constants and will be interpreted as the magnetic field and the chemical potential dual to electric charge. The function $f(r)$ obeys the following differential equation 
\be
rf\rq{}(r)+3f(r)+\Lambda R^2+\f{T_b R^2}{2r^2}\sqrt{r^4+R^4(\rho^2+\lambda^2 B^2)}=0
\ee

and  has the explicit form \cite{Pal:2012zn} 
\be
f(r)=\f{c_1}{r^3}-\f{\Lambda R^2}{3}- \f{T_bR^2}{6r^2}\sqrt{r^4+R^4(\rho^2+\lambda^2 B^2)}- \f{T_bR^4\sqrt{\rho^2+\lambda^2 B^2}}{3r^2}{}_2F_1\bigg[\f{1}{2},\f{1}{4},\f{5}{4},-\f{r^{4}}{R^4(\rho^2+\lambda^2 B^2)}\bigg]
\ee

In this case, the on-shell Einstein-Hilbert bulk action  gives
\be
S_b=\f{1}{2\kappa^2}\int d^{3+1}x\left[ \f{d}{dr}\left(-\f{\sqrt{g_{tt}}}{\sqrt{g_{rr}}}g\rq{}_{xx}\right)+T_b \f{\lambda^2B^2\sqrt{g_{tt}g_{rr}}}{\sqrt{\rho^2+\lambda^2B^2+g^2_{xx}}}\right]
\ee 

In what follows, we shall consider the form of the geometry  as written in eq(\ref{dyon_sol}). In which case, the bulk action reduces to
\be
S_b=\f{1}{2\kappa^2}\int d^{3}x\left(-2\f{r^3}{R^4}f(r)+T_b~r\f{\lambda^2 B^2}{\sqrt{\rho^2+\lambda^2 B^2}} {}_2F_1\bigg[\f{1}{2},\f{1}{4},\f{5}{4},-\f{r^{4}}{R^4(\rho^2+\lambda^2 B^2)}\bigg]\right)^{UV}_{IR}
\ee

The Gibbons-Hawking term and the counter term takes the following form
\bea
S_{GH}+S_{ct}&=&\f{1}{2\kappa^2}\int d^{3}x\left(\f{g\rq{}_{tt}g_{xx}}{\sqrt{g_{tt}g_{rr}}} +2g\rq{}_{xx}\f{\sqrt{g_{tt}}}{\sqrt{g_{rr}}}+\alpha  \sqrt{g_{tt}}g_{xx}\right)^{UV}\nn
&=&\f{1}{2\kappa^2}\int d^{3}x\left( \f{3r^3f(r)}{R^4}- \f{\Lambda r^3}{R^2}- \f{T_b r}{2R^2}\sqrt{r^4+R^4(\rho^2+\lambda^2 B^2)}+\alpha \f{r^3}{R^3}\sqrt{f(r)}\right)^{UV},
\eea

The total action, $S_{total}\equiv S_b+S_{GH}+S_{ct}$ gives
\bea
S_{total}&=&\f{1}{2\kappa^2}\int d^{3}x\Bigg[\left(\f{r^3f(r)}{R^4}- \f{\Lambda r^3}{R^2}- \f{T_b r}{2R^2}\sqrt{r^4+R^4(\rho^2+\lambda^2 B^2)}+\alpha \f{r^3}{R^3}\sqrt{f(r)}\right)^{UV}\nn
&+&T_b~\left(r\f{\lambda^2 B^2}{\sqrt{\rho^2+\lambda^2 B^2}} {}_2F_1\bigg[\f{1}{2},\f{1}{4},\f{5}{4},-\f{r^{4}}{R^4(\rho^2+\lambda^2 B^2)}\bigg]\right)^{UV}_{IR}\Bigg],
\eea
where we have considered, $f(r_{IR})=0$. It means for a black hole solution, $r_{IR}=r_h$.

Upon demanding that the  total  action, $S_{total}$,  is finite at the boundary allows us to set $\alpha=-\f{4}{R}\sqrt{\f{-2\Lambda_{eff}R^2}{6}}$ and in which case, it reduces to 
\be
S_{total}=-\f{1}{2\kappa^2}V_2\beta \Bigg[\f{c_1}{R^4}+T_b~r_h\f{\lambda^2 B^2}{\sqrt{\rho^2+\lambda^2 B^2}} {}_2F_1\bigg[\f{1}{2},\f{1}{4},\f{5}{4},-\f{r^{4}_h}{R^4(\rho^2+\lambda^2 B^2)}\bigg]\Bigg],
\ee
where $\beta$ is the periodicity of the Euclidean time circle. The quantity, $c_1$, can be determined from the condition, $f(r_h)=0$. Hence, we get  

\bea
S_{total}&=&-\f{V_2\beta}{2\kappa^2} \Bigg[\f{\Lambda r^3_h}{3R^2}+\f{T_br_h}{6R^2}\sqrt{r^4_h+R^4(\rho^2+\lambda^2 B^2)}+\f{T_br_h(\rho^2+4\lambda^2 B^2)}{3\sqrt{\rho^2+\lambda^2 B^2}}\times\nn&&{}_2F_1\bigg[\f{1}{2},\f{1}{4},\f{5}{4},-\f{r^{4}_h}{R^4(\rho^2+\lambda^2 B^2)}\bigg]\Bigg]\nn
&=&-\f{V_2\beta}{2\kappa^2} \Bigg[\f{\Lambda r^3_h}{3R^2}-\f{T_br_h\lambda^2 B^2}{3R^2(\rho^2+\lambda^2 B^2)}\sqrt{r^4_h+R^4(\rho^2+\lambda^2 B^2)}+\f{T_br_h(\rho^2+4\lambda^2 B^2)}{2\sqrt{\rho^2+\lambda^2 B^2}}\times\nn&&{}_2F_1\bigg[-\f{1}{2},\f{1}{4},\f{5}{4},-\f{r^{4}_h}{R^4(\rho^2+\lambda^2 B^2)}\bigg]\Bigg]\equiv -I,
\eea
where $I/\beta$ is the thermodynamic potential, $\Omega$. 
It is very easy to see that in the zero magnetic field limit the quantity $S_{total}$ reduces to that of the electrical charged black hole case studied earlier for $d=3$. The temperature, $T_H$,  and entropy, $S$, of such a dyonic back hole reads as
\be
T_H=-\f{r_h}{8\pi}\left[2\Lambda+\f{T_b}{r^2_h}\sqrt{r^4_h+R^4(\rho^2+\lambda^2 B^2)}\right],\quad s\equiv \f{S}{V_2}=\f{2\pi}{\kappa^2}\f{r^2_h}{R^2}.
\ee

\begin{figure}[h!]
\centering
   {\includegraphics[ width=8cm,height=6cm]{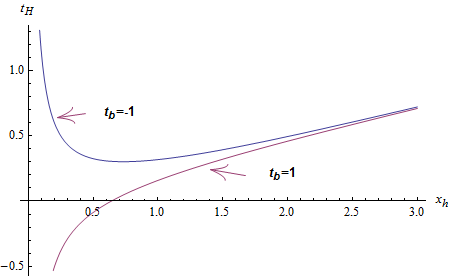} }
  \caption{
   The  figure is   plotted for the temperature vs the size of the horizon for  $AdS_4$ black hole.  The parameters are set as: ${\tilde \lambda}=-\f{d(d-1)}{2}-\f{t_b}{2},~d=3,~\rho=-2,~\lambda B=2$. For negative brane tension there exists two branches whereas for positive brane tension there exists only one branch.  }
 \label{fig_6}
\end{figure}

The behavior of the temperature versus the size of the horizon is plotted in fig(\ref{fig_6}) for fixed charge density and magnetic field. Again for negative tension of the brane, there exists two different branches. These two branches have two different signs of the specific heat. 

\paragraph{Null energy condition:} The imposition of the null energy condition, $T_{MN} u^M u^N \geq 0$, for null vectors, $u^M$ can be re-written as ${\cal R}_{MN} u^M u^N \geq 0$ upon using the solution  of the metric 
\be
{\cal R}_{MN} u^M u^N=\f{r^2f''(r)+4 r f'(r)}{2R^2}=\f{T_b}{2r^2} \f{R^4(\rho^2+\lambda^2 B^2)}{\sqrt{r^4+R^4(\rho^2+\lambda^2 B^2)}},
\ee
where we have used the null vector as given earlier in choice (b) in section (2). It just follows that the null energy condition imposes a restriction on the tension of the brane, $T_b$, that it should always be positive.

\begin{figure}[h!]
\centering
   {\includegraphics[ width=8cm,height=6cm]{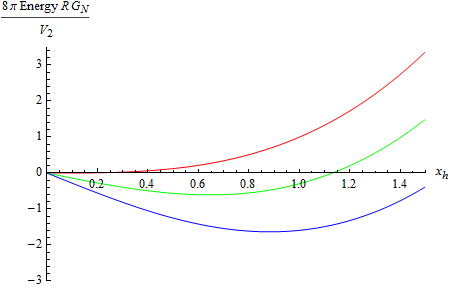} }
  \caption{
   The  figure is   plotted for the   energy density times $8\pi G_N R$ vs the  horizon size for the  $AdS_4$ black hole.  The parameters are set as: ${\tilde \lambda}=-\f{d(d-1)}{2}-\f{t_b}{2},~\rho=-2,~\lambda B=2$. The red curve, green curve and the blue curve are for $t_b=0,~~t_b=1$ and $t_b=2$, respectively.}
\label{fig_7}
\end{figure}

\paragraph{Energy:} The energy of a dyonic black hole can be calculated using the formula as written in eq(\ref{energy}). It reads as
\be\label{energy_dyonic}
E=-\f{1}{8\pi G}\f{V_{2}}{R^{4}} \left(\f{\Lambda R^2 r^3_h}{3} + \f{T_b r_h R^4\sqrt{\rho^2+\lambda^2 B^2}}{2}{}_2F_1\bigg[-\f{1}{2},\f{1}{4},\f{5}{4},-\f{r^{4}_h}{R^4(\rho^2+\lambda^2 B^2)}\bigg] \right),
\ee
which is plotted in fig(\ref{fig_7})

\paragraph{First law of thermodynamics:} Given the energy, $E$, as written above in eq(\ref{energy_dyonic}),  the variation of the energy, $E$,  gives
\bea
dE&=&\f{1}{8\pi G}\f{V_{2}}{R^{4}} \Bigg( 4\pi T_H R^2 r_h dr_h-\f{T_b R^4\rho r_h}{2\sqrt{\rho^2+\lambda^2B^2}}{}_2F_1\bigg[\f{1}{2},\f{1}{4},\f{5}{4},-\f{r^{4}_h}{R^4(\rho^2+\lambda^2 B^2)}\bigg] d\rho\nn&-&\f{T_b R^4\lambda^2B r_h}{2\sqrt{\rho^2+\lambda^2B^2}}{}_2F_1\bigg[\f{1}{2},\f{1}{4},\f{5}{4},-\f{r^{4}_h}{R^4(\rho^2+\lambda^2 B^2)}\bigg] dB\Bigg)\nn
&=&T_H~ dS+\varphi_E ~dq_E-\f{T_b R^4\lambda^2B r_h}{2\sqrt{\rho^2+\lambda^2B^2}}{}_2F_1\bigg[\f{1}{2},\f{1}{4},\f{5}{4},-\f{r^{4}_h}{R^4(\rho^2+\lambda^2 B^2)}\bigg] dB,
\eea
 where the entropy, $S$, chemical potential, $\varphi_E$ and the electric charge, $q_E$ are
 \be
 S=\f{V_2 r^2_h}{4G R^2},\quad \varphi_E=-\f{r_h \rho}{\lambda\sqrt{\rho^2+\lambda^2 B^2}}{}_2F_1\bigg[\f{1}{2},\f{1}{4},\f{5}{4},-\f{r^{4}_h}{R^4(\rho^2+\lambda^2 B^2)}\bigg],\quad q_E=\f{T_b V_2 \lambda}{16\pi G}\rho
 \ee

If we write the magnetic charge and the chemical potential associated to it as
\be
q_M=\f{T_b V_2 \lambda^2}{16\pi G}  B,\quad \varphi_M=-\f{r_h B}{\sqrt{\rho^2+\lambda^2 B^2}}{}_2F_1\bigg[\f{1}{2},\f{1}{4},\f{5}{4},-\f{r^{4}_h}{R^4(\rho^2+\lambda^2 B^2)}\bigg]
\ee

then the variation of the energy can be written as 
\be
dE=T_H~ dS+\varphi_E ~dq_E+\varphi_M ~dq_M.
\ee

Generically, the magnetic charge is defined as
\be
q_M=\f{T_b  \lambda^2}{16\pi G}\int F,
\ee
where $F$ is a two form defined in eq(\ref{dyon_sol}) and $\int F$ is to be evaluated in the large $r$ limit. Note that $lim_{r\ra\infty}A_t(r)\ra\varphi_E$.

\paragraph{Smarr Relation:} The relation between  energy,  temperature, electric charge, electric type chemical potential, magnetic charge and magnetic type chemical potential is calculated to have the following form 
\be
E=\f{2}{3}\left[ T_H S+q_E\varphi_E+q_M \varphi_M\right]
\ee

\paragraph{Ratio $\f{C_{\rho}}{s}$:} Upon doing calculations, we find  the ratio comes out as 
\be
\f{C_{\rho}}{s}=2\left[1+\f{T_b R^4(\rho^2+\lambda^2 B^2)}{4\pi T_H r_h\sqrt{r^4_h+R^4(\rho^2+\lambda^2 B^2)}}\right]^{-1}
\ee

It has the same feature as obtained for the charged black hole case, namely, for large black holes ($r_h\ra\infty$), the ratio $\f{C_{\rho}}{s}$ approaches $2$. It is thermodynamically stable for  positive tension of the brane.

\section{Conclusion}

In this paper, we have studied the thermodynamics of the Einstein-Dirac-Born-Infeld (EDBI) action. 
The thermodynamics is studied by calculating the free energy of the EDBI system in the grand canonical ensemble. In particular, the free energy is computed by regulating the on-shell bulk action with the help of the Gibbon-Hawking term along with a counter term. The coefficient of  the counter term is fixed so as to remove the infinity at the boundary, i.e., at UV, of the on-shell action.  Once the on-shell action is regularized various thermodynamic quantities are computed from the free energy using the usual thermodynamic prescription.

In the study of the thermodynamics, we have shown that the Einstein-Dirac-Born-Infeld system obeys the first law of  black hole.  Moreover,  it also respects the Smarr formula. It is thermodynamically stable. Interestingly, the first law of thermodynamics does not depend on the validity of the null energy condition.

Let us make a comprehensive list of  comparison of the thermodynamical quantities that follows from the Einstein-Maxwell system  and the Einstein-Dirac-Born-Infeld system in $3+1$ spacetime dimensions. 

\paragraph{Grand Potential:} The precise form of the grandpotential  in  various powers of the size of the horizon for the dyonic black hole in $3+1$ dimensional spacetime dimensions are 
\bea
\Omega_{Maxwell}&=&a ~~r^3_h+\f{b}{r_h},\quad \textnormal{ where\quad a, b \quad are constants} \nn 
\Omega_{DBI}&=& \left\{
\begin{array}{lr}
a_1 r^3_h+\f{b_1}{r_h}+\f{c_1}{r^5_h}+\cdots &  \frac{(\rho^2+\lambda^2 B^2)R^4}{r^4_h} \ll 1\\
a_2 r_h+b_2r^3_h+c_2r^5_h+\cdots &  \frac{(\rho^2+\lambda^2 B^2)R^4}{r^4_h} \gg 1
\end{array}
\right.
\eea
for some constants $a_i,~b_i$ and $c_i$. 
\paragraph{Temperature:} The Hawking temperature reads as
\bea
T_{Maxwell}&=&A ~~r_h+\f{B}{r^3_h},\quad \textnormal{ where\quad A,  B \qquad are constants} \nn 
T_{DBI}&=& \left\{
\begin{array}{lr}
A_1 r_h+\f{B_1}{r^3_h}+\f{C_1}{r^7_h}+\cdots &  \frac{(\rho^2+\lambda^2 B^2)R^4}{r^4_h} \ll 1\\
\f{A_2}{ r_h}+B_2r_h+C_2r^3_h+\cdots &  \frac{(\rho^2+\lambda^2 B^2)R^4}{r^4_h} \gg 1
\end{array}
\right.
\eea
for some constants $A_i,~B_i$ and $C_i$. 
\paragraph{Chemical potential:}
\bea
\varphi_{E,~Maxwell}&=&\f{x}{r_h},\qquad \varphi_{M,~Maxwell}=\f{y}{r_h},\quad \textnormal{ where\quad x,y \quad are constants} \nn 
\varphi_{E,~ DBI}&=& \left\{
\begin{array}{lr}
\f{X_1}{ r_h}+\f{Y_1}{r^5_h}+\cdots &  \frac{(\rho^2+\lambda^2 B^2)R^4}{r^4_h} \ll 1\\
X_2 r_h+Y_2r^5_h+\cdots &  \frac{(\rho^2+\lambda^2 B^2)R^4}{r^4_h} \gg 1,
\end{array}
\right.
\eea
where $X_i,~Y_i$'s are constants. 
Similar structure for the magnetic part of the chemical potential for the Einstein-DBI system. 

\paragraph{Energy:} The energy associated to the Einstein-Maxwell black hole and the Einstein-DBI balck hole are 
\bea
E_{Maxwell}&=&s ~~r^3_h+\f{\tau}{r_h},\quad \textnormal{ where\quad s and  $\tau
	$ \quad are constants} \nn 
E_{DBI}&=& \left\{
\begin{array}{lr}
	s_1 r^3_h+\f{\tau_1}{r_h}+\f{\xi_1}{r^5_h}+\cdots &  \frac{(\rho^2+\lambda^2 B^2)R^4}{r^4_h} \ll 1\\
	s_2 r_h+\tau_2r^3_h+\xi_2r^5_h+\cdots &  \frac{(\rho^2+\lambda^2 B^2)R^4}{r^4_h} \gg 1,
\end{array}
\right.
\eea
for some constant $s_i,~\tau_i$ and $\xi_i$'s.

\paragraph{Entropy:} The form of the  entropy in terms of the size of the horizon is same for both the systems. However, the size of the horizon for both the cases are not necessarily same. In the Einstein-Dirac-Born-Infeld case, it depends on the tension on the tension of brane, $T_b$ as well as on the background cosmological constant. However, for the Einstein-Maxwell case, it  depends on the cosmological constant. The dependence on the electric charge density and the magnetic field differs in both the cases.

It is interesting to note that all the thermodynamical quantities, except the entropy, are  odd powers in the size of the horizon. Moreover, in the limit of $\frac{(\rho^2+\lambda^2 B^2)R^4}{r^4_h} \ll 1$, the Einstein-DBI system gives the same result as Einstein-Maxwell system. It means in the limit of small electric charge density or the magnetic field in comparison to the size of the horizon the Einstein-DBI system gives the same physics as that of the Einstein-Maxwell system.

Various thermodynamical quantities for large size of the horizon provides the known AdS result, as expected. Different kinds of matter field gives rise to result which are sub-leading to the AdS result at UV, whereas for small size of the horizon i.e., at IR it can give different result.


 \paragraph{ Acknowledgment:} It is a pleasure to thank the anonymous referee for suggesting to check the energy condition. Thanks are to  arxiv.org and the people behind it for their unconditional support to help users like me to pursue their research, which otherwise simply won\rq{}t have been possible. 
 
\section{Appendix: A }

In this appendix, we shall collect some properties of the hypergeometric function, ${}_2F_1[a,b,c,x]\equiv F[a,b,c,x]$ from \cite{as}.

\bea
F[a,b,c,-x]&=&(1+x)^{-a}F[a,c-b,c,\f{x}{1+x}]
\eea


In the limit of large, $x\ra\infty$, the leading order term is 
\bea
lim_{x\ra\infty}F[a,b,c,-x]&\ra& x^{-a}\left(1-\f{a}{x}+\cdots\right) F[a,c-b,c,1]\nn
&=&x^{-a}\f{\Gamma\left(c\right)\Gamma\left(b-a\right)}{\Gamma\left(c-a\right)\Gamma\left(b\right)}+{\cal O}(\f{1}{x^{a+1}})
\eea


In particular, for $a=-1/2, ~b=1/(2(d-1)),~ c=1+b, ~x=1/\rho^2~ (r/R)^{2(d-1)}, $ the leading order term for large $r$ is 
\bea
&&lim_{r\ra\infty}F\left[-\f{1}{2},\f{1}{2(d-1)},1+\f{1}{2(d-1)},-\f{1}{\rho^2}~\left (\f{r}{R}\right)^{2(d-1)}\right]= \f{1}{\rho}(r/R)^{d-1}\times \f{1}{d}+{\cal O}(1/r^{d-1})\nn
\eea
where we have used $,F\left[-\f{1}{2},1,1+\f{1}{2(d-1)},~1\right]=\f{1}{d}$.


\paragraph {Electrically charged black hole:} In this case, the  function, $f(r)$, has the form 
\be
f(r)=\f{c_1}{r^d}-\f{2\Lambda R^2}{d(d-1)}-\f{T_b R^2\rho}{(d-1)}\f{r^{1-d}}{R^{1-d}}~~~{}_2F_1\bigg[-\f{1}{2},\f{1}{2(d-1)},\f{2d-1}{2(d-1)},-\f{r^{2(d-1)}}{R^{2(d-1)}\rho^2}\bigg]. 
\ee
whose asymptotic behavior i.e.,  at UV is 
\bea
&&lim_{r\ra\infty}~f(r)= \f{c_1}{r^d}-\f{2\Lambda_{eff} R^2}{d(d-1)}+{\cal O}\left(\f{1}{r^{2(d-1)}}\right),\nn
&&lim_{r\ra\infty}~\sqrt{f(r)}=\sqrt{-\f{2\Lambda_{eff} R^2}{d(d-1)}}\left[1-\f{c_1}{2r^d}\f{d(d-1)}{2\Lambda_{eff} R^2}\right]+{\cal O}\left(\f{1}{r^{2(d-1)}}\right),\nn
&&lim_{r\ra\infty}~\left((d-2)f(r)+\alpha R\sqrt{f(r)}\right)=\sqrt{-\f{2\Lambda_{eff} R^2}{d(d-1)}}\left[\alpha R+(d-2)\sqrt{-\f{2\Lambda_{eff} R^2}{d(d-1)}}\right]\nn&+&
\f{c_1}{r^d}\left[(d-2)+\f{\alpha R}{2}\sqrt{\f{d(d-1)}{-2\Lambda_{eff} R^2}} \right]+{\cal O}\left(\f{1}{r^{2(d-1)}}\right)\nn
&&lim_{r\ra\infty}~\left[-\f{2\Lambda R^2}{d-1}-\f{T_bR^2}{d-1} \f{r^{1-d}}{R^{1-d}}\sqrt{\rho^2+\f{r^{2(d-1)}}{R^{2(d-1)}}}+(d-2)f(r)+\alpha R\sqrt{f(r)}\right]=\nn&&-\f{4\Lambda_{eff}R^2}{d}
+\alpha R \sqrt{-\f{2\Lambda_{eff} R^2}{d(d-1)}}+\f{c_1}{r^d}\left[(d-2)+\f{\alpha R}{2}\sqrt{\f{d(d-1)}{-2\Lambda_{eff} R^2}} \right]+{\cal O}\left(\f{1}{r^{2(d-1)}}\right)\nn
\eea
where $2\Lambda_{eff}=T_b+2\Lambda$.

Now, because of the brane, which is described by DBI action,  the cosmological constant is given by $\Lambda_{eff}$. In this case, we can re-define $R$ and the spatial coordinates, $x_i$, to bring the geometry to AdS at UV.

\paragraph{Dyonic black hole:} For such a black hole, the function, $f(r)$,  is given in \cite{Pal:2012zn} and has the form 
\bea
f(r)&=&\f{c_1}{r^3}-\f{\Lambda R^2}{3}- \f{T_bR^2\sqrt{r^4+R^4(\rho^2+\lambda^2 B^2)}}{6r^2}- \f{T_bR^4\sqrt{\rho^2+\lambda^2 B^2}}{3r^2}{}_2F_1\bigg[\f{1}{2},\f{1}{4},\f{5}{4},-\f{r^{4}}{R^4(\rho^2+\lambda^2 B^2)}\bigg],\nn
&=&\f{c_1}{r^3}-\f{\Lambda R^2}{3}- \f{T_bR^4\sqrt{\rho^2+\lambda^2 B^2}}{2r^2}{}_2F_1\bigg[-\f{1}{2},\f{1}{4},\f{5}{4},-\f{r^{4}}{R^4(\rho^2+\lambda^2 B^2)}\bigg]
\eea

At UV, the function, $f(r)$ behaves as
\bea
&&lim_{r\ra\infty}~f(r)=-\f{2\Lambda_{eff} R^2}{6}+\f{c_1}{r^3}+{\cal O}\left(\f{1}{r^4}\right),\nn
&&lim_{r\ra\infty}~\sqrt{f(r)}=\sqrt{-\f{2\Lambda_{eff} R^2}{6}}+\f{c_1}{2r^3}\sqrt{\f{6}{-2\Lambda_{eff} R^2}}+{\cal O}\left(\f{1}{r^{4}}\right),\nn
\eea

\section{Appendix B: Einstein-Maxwell system  (Dyonic AdS black hole)}

In this section, we shall calculate the free energy associated to the dyonic  AdS black hole. For this purpose, we shall consider the following action
\be\label{ads_bulk_action}
S_b=\f{1}{2\kappa^2}\int \sqrt{-g}\left( {\cal R}-R^2 F_{MN}F^{MN}+\f{6}{R^2}\right)
\ee

The solution takes the following form
\bea\label{dyon_sol_maxwell}
ds^2_{d+1}&=&\f{r^2}{R^2}\left[-f(r)dt^2+dx^2+dy^2\right]+\f{R^2 dr^2}{r^2 f(r)},\quad F=A\rq{}_t(r)dt\w dr+b dx\w dy\nn
A_t(r)&=&\varphi_E-\f{\rho_E R^2}{r},
\eea
where  $\varphi_E=\f{\rho_E R^2}{r_h}$  will be interpreted as  the chemical potential dual to electric charge and $\rho_E$ is the charge density. The function $f(r)$ obeys the following differential equation 
\be
rf\rq{}(r)+3f(r)-3+R^8\f{(\rho^2_E+ b^2)}{r^4}=0
\ee

and the has the explicit form as 
\be
f(r)=1+\f{c_1}{r^3}+R^8\f{(\rho^2_E+ b^2)}{r^4}.
\ee

The finite on-shell value of the action, which is the sum of the bulk action, eq(\ref{ads_bulk_action}),  along with the Gibbons-Hawking term and the counter term gives
\be
S_{total}=S_b+S_{GH}+S_{ct}=-\f{V_2\beta}{2\kappa^2}\left[ \f{c_1}{R^4}+\f{4R^4b^2}{r_h}\right]\equiv - I,
\ee
where $S_{GH},~S_{ct}$ are written in eq({\ref{gh_action}}) and eq(\ref{ct_action}) and $r_h$ is the horizon for which $f(r_h)=0$. The only value of $\alpha$, for which $S_{total}$ becomes finite is  $\alpha=-4/R$.  In which case, the grand potential becomes
\be
\Omega\equiv \f{I}{\beta}=-\f{V_2}{2\kappa^2}\left[\f{r^3_h}{R^4} +\f{(\rho^2_E-3b^2)}{r_h}R^4\right]=-\f{V_2}{2\kappa^2}\left[\f{r^3_h}{R^4}+\varphi^2_E r_h-\f{3b^2}{r_h}R^4\right].
\ee

The Hawking temperature,$T_H$,  and the Hawking-Bekenstein entropy, $S$,  of such a dyonic black hole reads as
\be
T_H=\f{3}{4\pi}\f{r_h}{R^2}\left( 1-\f{(\rho^2_E+ b^2)}{3r^4_h}R^8\right),\quad S=\f{V_2}{4G}\left(\f{r_h}{R}\right)^2
\ee

Given the temperature as written above, one can very easily see that there exists only one branch for which the temperature is positive and this happens for large size of the horizon. Moreover, this branch has positive specific heat and is thermodynamically stable.

Defining the dimensional variable as $x_h\equiv r_h/R,~~~t_H\equiv T_H R,~~~{\tilde \rho}_E\equiv\rho_E R^2,~~~{\tilde b}\equiv bR^2$, we can write the temperature as 
\be
t_H=\f{3}{4\pi}x_h\left(1-\f{({\tilde \rho}^2_E+{\tilde b}^2)}{3x^4_h}\right) 
\ee

\begin{figure}[h!]
\centering
   {\includegraphics[ width=8cm,height=6cm]{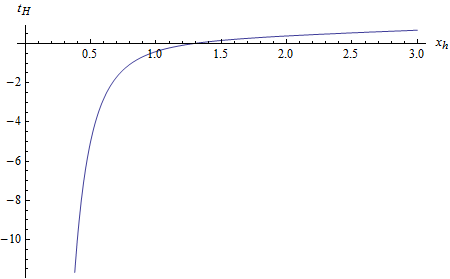} }
  \caption{
   The  figure is   plotted for the temperature vs the size of the horizon for  $AdS_4$ black hole.  The parameters are set as: ${\tilde \rho}_E=-2,~{\tilde b}=2$.There exists  only one branch.  }
 \label{fig_8}
\end{figure}

The dimensionless temperature is plotted in fig(\ref{fig_8}), from which it follows that there exists only one branch.

\paragraph{Energy:} The energy that follows from eq(\ref{energy}) for $d=3$ is
\be
E=-\f{V_2}{8\pi G} \f{c_1}{R^4}=\f{V_2}{8\pi G}\left[\f{r^3_h}{R^4}+\f{(\rho^2_E+b^2)R^4}{r_h}\right]
\ee

\begin{figure}[h!]
\centering
   {\includegraphics[ width=8cm,height=6cm]{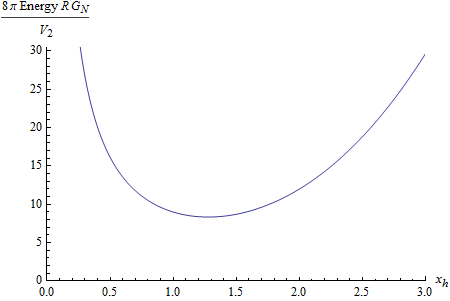} }
  \caption{
   The  figure is   plotted for the   energy density times $8\pi G_N R$ vs the  horizon size for the  $AdS_4$ black hole.  The parameters are set as: ${\tilde \rho}_E=-2,~{\tilde b}=2$.}
\label{fig_9}
\end{figure} 

Using the above mentioned dimensionless variable, the figure for the energy is plotted in fig(\ref{fig_9}).
\paragraph{First law of thermodynamics:} The variation of the energy gives 
\be
dE= T_H ~dS+\varphi_E \f{V_2R^2}{4\pi G}d\rho_E+\f{V_2R^2}{4\pi G}\left(\f{bR^2}{r_h}\right)db
\ee

Recall, that the electric charge, $q_E=-\f{V_2}{g^2_{YM}} \sqrt{-g} F^{r0}$. In our case $g^2_{YM}=\f{4\pi G}{R^2}$, which gives the electric charge 
\be
q_E=\f{V_2R^2}{4\pi G}\rho_E.
\ee

Given the field strength tensor $F=A\rq{}_t(r)dt\w dr+b dx\w dy$, we can find the dual field strength ${}^{\star}F=-\f{r^2}{R^2}A\rq{}_t(r) dx\w dy-b\f{R^2}{r^2} dr\w dt$. It follows  that $d{}^{\star}F=0$, upon using the equation of motion associated to 
$A_t(r)$, which suggests us to write ${}^{\star}F=d{\cal A}$, where ${\cal A}$ is a one-form potential. The electric component of  ${\cal A}$ gives
\be
{\cal A}_t=b\f{R^2}{r}+\varphi_M,
\ee
where $\varphi_M$ is a constant and has to obtained in such a way that the norm of  ${\cal A}$ vanishes at the horizon. It gives
\be
\varphi_M=-b\f{R^2}{r_h},
\ee
which will be interpreted as the magnetic part of the chemical potential \cite{Lu:2013ura}.

If we want the first law of thermodynamics to be of the following form: $dE=T_H dS+\varphi_E dq_E+\varphi_M dq_M$, then it just follows that the magnetic charge, $q_M=-\f{V_2R^2}{4\pi G}b$. This essentially means a suggestive form of the magnetic form of the conserved charge is $q_M=\f{V_2}{g^2_{YM}} \sqrt{-g}~   {}^{\star}F^{r0}$.

Now, we can re-express the grand potential, $\Omega(r_h,~\varphi_E,~q_M)$, as
\be
\Omega=-\f{V_2}{2\kappa^2}\left[\f{r^3_h}{R^4}+\varphi^2_E r_h-\f{48\pi^2G^2}{r_h}\f{q^2_M}{V^2_2}\right].
\ee

 By considering the variation of the grand potential as
\be
d\Omega=-S~ dT_H-q_E~ d\varphi_E +\varphi_M ~dq_M,
\ee

we can read out the entropy, $S$, electric charge, $q_E$, and the magnetic part of the chemical potential, $\varphi_M$, which in fact matches with the result reported above.

\paragraph{Ratio $\f{C_{\rho}}{s}$:} Upon doing calculations, we find  the ratio comes out as 
\be
\f{C_{\rho}}{s}=2\left[1+\f{ R^6(\rho^2_E+b^2)}{\pi T_H r^3_h}\right]^{-1}
\ee

Again for large size of the black hole horizon, it approaches two. However, for smaller size of the horizon, it never becomes negative, because the size of the horizon cannot go below a particular value as dictated by the positive value of the Hawking temperature.

\section{Appendix C}

In this appendix, we shall investigate whether its possible to write the dual of the field strengh tensor as ${}^{\star} F=d{\cal A}$, for a one-form ${\cal A}$.

For our purpose let the field strength tensor is described by $F=A\rq{}_t(r)dt\w dr+b dx\w dy$, in the  four dimensional black hole AdS spacetime, where $b$ is a constant quantity. Let us also assume that the geometry has the following explicit form
\be
ds^2=\f{r^2}{R^2}\left[-f(r)dt^2+dx^2+dy^2\right]+\f{R^2 dr^2}{r^2 f(r)}
\ee

 then    the dual field strength ${}^{\star}F=-\f{r^2}{R^2}A\rq{}_t(r) dx\w dy-b\f{R^2}{r^2} dr\w dt$. The question that we ask can we write  ${}^{\star} F=d{\cal A}$, for some one-form ${\cal A}$? 
 
 Let us assume that it is possible to write ${}^{\star} F=d{\cal A}$. If so then $d {}^{\star} F=0$. It means   ${}^{\star} F=d{\cal A}$  amounts to check whether  $d {}^{\star} F=0$ or not.
 
 Upon calculating  $d {}^{\star} F=-\partial_r\left[\f{r^2}{R^2}A\rq{}_t(r) \right]
 dr\w dx\w dy$, so  after imposing  $d {}^{\star} F=0$ we get the relation that need to be checked 
 $\partial_r\left[\f{r^2}{R^2}A\rq{}_t(r) \right]=0$.
 
 \paragraph{Maxwell case:} In this case the equation of motion for the gauge field is $\partial_M \left[\sqrt{-g}F^{MN} \right]=0$, which in the differential form language can be written  as $d {}^{\star} F=0$.
So, the time component of the gauge potential, ${\cal A}_t(r)$, as defined in ${}^{\star} F=d{\cal A}$ reads as ${\cal A}_t(r)=b\f{R^2}{r}+{\rm constants}.$ It means the magnetic part of the chemical potential reads as $\varphi_M=-b\f{R^2}{r_h}.$

\paragraph{DBI case:} In this case the equation of the gauge field reads as \\
$\partial_M \left[\sqrt{-det(g+\lambda F)}\left((g+\lambda F)^{-1}-(g-\lambda F)^{-1}\right)^{MN} \right]=0$. This can be further reduced to $\partial_r \left[\lambda A'_t\f{\sqrt{r^4+R^4\lambda^2b^2}}{\sqrt{1-\lambda^2 A'^2_t}}\right]=0$.  So  in this case it is not a priori clear 
how to read out the magnetic part of the chemical potential from the field strength. However, if we demand the validity of the first law of thermodynamics, $dE= T_H dS+\varphi_E dq_E+\varphi_M dq_M$, then we can read out very easily the magnetic part of the chemical potential, $\varphi_M$.

\end{document}